\documentclass[aps,prd,groupedaddress,floatfix,nofootinbib]{revtex4}
\usepackage{amsmath}
\usepackage{graphicx}
\usepackage{color}
\usepackage{graphicx}
\usepackage{epsfig}
\usepackage{amsfonts}
\usepackage{amstext}
\usepackage{amssymb}
\newcommand{\beq}{\begin{eqnarray}}
\newcommand{\eeq}{\end{eqnarray}}

\begin{document}

\title{Failure of intuition in elementary rigid body dynamics}
\author{Nivaldo A. Lemos}
\email{nivaldo@if.uff.br}
\affiliation{Instituto de F\'{\i}sica, Universidade Federal Fluminense,\\
Av. Litor\^anea  s/n$^o$,   24210-340, Niter\'oi, RJ, Brazil}

\begin{abstract}
Suppose a projectile collides perpendicularly with a stationary rigid rod on a smooth horizontal table. We show that, contrary to what one naturally expects, it is not always the case that the rod acquires maximum angular velocity when struck at an extremity. The treatment is intended for
first year university students of Physics or Engineering, and could form the basis of a tutorial discussion of conservation laws in rigid body dynamics.
\end{abstract}

\maketitle


On top of being a mathematically difficult subject, the general rotational dynamics  of rigid bodies often defies our intuitive expectations. Although in elementary cases physical intuition usually works, here we wish to discuss a very simple problem in rigid body dynamics with at least one aspect that does not conform to what intuition suggests.

In the collision shown in Figure 1, it is natural to expect that bigger $a$ gives bigger $\omega$, with $a = l$ giving the largest $\omega$. Our everyday experience with doors and levers  indicates that the greatest rotational motion is effected by applying a force  as far as possible from the rotation axis or from the fulcrum. If the rod is hit dead centre, its resulting motion is translational only, and as $a$ increases the angular momentum is expected to increase.  However, we will show that if the rod has low mass compared to the projectile the maximum angular velocity is obtained with $a < l$, which is surprising.

What we undertake  to examine in detail is a well-known textbook problem \cite{Resnick}. On a frictionless horizontal table a thin homogeneous rod, of mass $M$ and length $2l$, at rest, is hit by a projectile of mass $m$ and speed $v_0$. The rod is struck perpendicularly at a distance $a$ from its centre. Assuming the collision is elastic, the problem is, of course, to find out how projectile and rod move after the collision.

\vspace{.5cm}
\begin{figure}[h!]
\epsfxsize=9cm
\begin{center}
\leavevmode
\epsffile{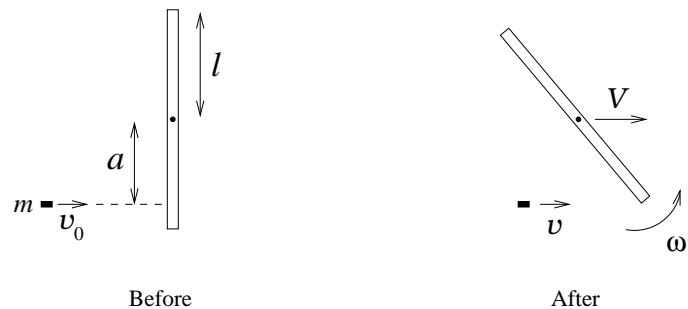}
\caption{A projectile collides elastically with a rigid rod at rest on a smooth horizontal table.}
\end{center}
\end{figure}
\vspace{-.2cm}
\vspace{.5cm}

Since there are no external forces or torques, linear momentum and angular momentum are conserved. Taking angular momenta with respect to the centre of the rod before the collision (a fixed point in an inertial frame), conservation of linear momentum, angular momentum and kinetic energy yield,  respectively:

\begin{equation}
\label{linear}
mv_0=mv + MV \, ;
\end{equation}

\begin{equation}
\label{angular}
mv_0a=mva + I\omega \, ;
\end{equation}

\begin{equation}
\label{kinetic}
\frac{m}{2}v_0^2=\frac{m}{2}v^2 + \frac{M}{2}V^2 + \frac{I}{2} \omega^2\, .
\end{equation}

\noindent In the above equations $v$ and $V$ are,  respectively,   the speeds of the projectile and of the rod's centre of mass after the collision, $\omega$ is the magnitude of the angular velocity acquired by the rod about    its centre of mass and $I$ is the moment of inertia of the rod with respect to a perpendicular axis through its centre of mass.

From (\ref{linear}) and (\ref{angular}) we have

\begin{equation}
\label{linear-angular}
V = \frac{m}{M}(v_0-v) \,\, ,\,\,\,\,\, \omega =\frac{ma}{I}\,(v_0 -v) \, ,
\end{equation}
with the use of which  Eq.(\ref{kinetic}) becomes

\begin{equation}
\label{kinetic2}
m(v_0^2 - v^2) = \frac{m^2}{M}(v_0 - v)^2 + \frac{m^2a^2}{I} (v_0 - v)^2\, .
\end{equation}
The solution $v=v_0$ is physically uninteresting, since it would mean no collision at all. Factoring out $\,v-v_0\,$ from Eq.(\ref{kinetic2})  leads to

\begin{equation}
\label{kinetic-divided}
m(v_0 + v) = \frac{m^2}{M}(v_0 - v) + \frac{m^2a^2}{I} (v_0 - v)\, ,
\end{equation}
which gives

\begin{equation}
\label{v}
 v = - \frac{\displaystyle 1- \frac{m}{M} -\frac{ma^2}{I} }{\displaystyle 1 + \frac{m}{M}  + \frac{ma^2}{I}}\, v_0 \, .
\end{equation}
The remaining quantities we are looking for are obtained by combining Eqs.(\ref{v}) and (\ref{linear-angular}):

\begin{equation}
\label{V}
 V =  \frac{\displaystyle 2m/M }{ \displaystyle 1 + \frac{m}{M}  + \frac{ma^2}{I} }\, v_0 \, ;
\end{equation}

\begin{equation}
\label{omega}
\omega =  \frac{\displaystyle 2 ma/I }{\displaystyle 1 + \frac{m}{M}  + \frac{ma^2}{I}}\, v_0 \, .
\end{equation}

These results reduce to what they should in simple particular or limiting cases: (i) if $a=0$ then $\omega =0$; (ii) if $a=0$ and $m=M$ it follows that  $v=0$ and $V=v_0$, the well-known result for a head-on elastic collision of two  bodies with the same mass; (iii) in the limit $M\to\infty$ (consequently $I\to \infty$ also) one has $v=-v_0$, $V=0$ and $\omega =0$, so that the projectile is reflected as from a wall.

Now we ask ourselves for what value of the impact parameter $a$ the rod's  angular velocity is greatest. Physical intuition strongly suggests that this occurs for $a=l$ whatever the masses.
It is an easy exercise to show that the function $\,f(x)=x/(b^2+c^2 x^2)\,$ reaches its  maximum at $\,x=b/c\,$ for positive $x$. Accordingly, from Eq.(\ref{omega}) it follows that the maximum angular velocity seems to be attained at

\begin{equation}
\label{maximal-a-general}
a= \sqrt{\frac{(M+m)I}{mM}}\, .
\end{equation}
For the homogeneous rod 

\begin{equation}
\label{inertia-homogeneous}
I= \frac{M (2l)^2}{12}= \frac{M l^2}{3} \, ,
\end{equation}
and Eq.(\ref{maximal-a-general}) yields
\begin{equation}
\label{maximal-a-homogeneous}
a= \sqrt{\frac{M+m}{3m}}\,\, l\, .
\end{equation}

Since the physical domain of values for $a$ is $\, 0\leq a \leq l$, if $M>2m$ the impact parameter furnished by Eq.(\ref{maximal-a-homogeneous}) is unphysical. In this case  $\omega$ is a monotonically increasing function of $a$ on the physical domain, so that the highest angular velocity is indeed attained if $a=l$. However, if $M\leq 2m$ the value given by (\ref{maximal-a-homogeneous}) is physically acceptable. If $M=2m$ one gets $a=l$, but for $M<2m$ the largest angular velocity is reached for some $a<l$. For $M=m$, for instance, $a= \sqrt{2/3}\, l$. If the rod is very light, $M\ll m$, the impact parameter that gives rise to the greatest angular velocity  may be nearly as small as $l/\sqrt{3}$. Surprises are not over yet. Exactly the same behavior is found
for a totally inelastic collision, case in which the projectile sticks to the rod. The analysis of the totally inelastic collision requires some additional care since the moment of inertia and the position of the centre of mass   of the rod are modified  after the collision because of the attached projectile.

Our daily observation of the behaviour of doors and levers is limited to rotational motion about a {\it fixed} axis, in which case the farther from the axis one applies a force the bigger the resulting angular velocity. In the case under discussion the rotation axis is not fixed, and it appears that intuition is misled by the fact that a {\it given} impulsive force produces the largest angular velocity if the force is applied at $a=l$.  However, the impulsive force acting during the  collision
 depends on the impact
parameter, as can be ascertained from the linear impulse delivered to the rod by the projectile. Indeed, if $P$ denotes the rod's linear momentum, it follows from Eq.(\ref{V}) that the linear impulse delivered to the rod is

\begin{equation}
\label{impulse-linear}
I_{lin}= \Delta P = MV=  \frac{\displaystyle 2m v_0}{ \displaystyle 1 + \frac{m}{M}  +  \frac{3ma^2}{Ml^2} }  \,,
\end{equation}
which gets smaller  as $a$ gets larger.
 Therefore, the angular  impulse
transmitted to the rod is
\begin{equation}
\label{impulse-angular}
I_{ang}= \frac{Ml^2}{3}\omega = a \Delta P = Ma V\, ,
\end{equation}
in which there is is a competition between an increasing $a$ and a decreasing $V$. As Eq.(\ref{impulse-linear}) shows, for  $M$ sufficiently large $V$ decreases slowly with increasing $a$, the impact parameter wins the competition and the maximum angular velocity is obtained with $a=l$.

It is instructive to consider the more general case of an inhomogeneous rod. For simplicicity, we assume that the mass distribution of the rod is symmetric with respect to its centre, so that the rod's middle point is still its  centre of mass. Writing

\begin{equation}
\label{inertia-general}
I= M r_{_G}^2 \, ,
\end{equation}
where $r_{_G}$ is the radius of gyration, Eq.(\ref{maximal-a-general}) yields
\begin{equation}
\label{maximal-a-gyration}
a= \sqrt{\frac{M+m}{m}}\,r_{_G}\, .
\end{equation}
If the mass distribution is much more concentrated near the middle of the rod, $r_{_G} \ll l$ and one finds $a < l$ for the impact parameter that gives the largest angular velocity even for $M\gg m$. If the mass distribution is highly concentrated close to the ends of the rod, $r_{_G}\lesssim l$ and only for $M\ll m$ would Eq.(\ref{maximal-a-gyration})  furnish $a<l$. It is curious that intuition is fully vindicated only in the case of an ideal dumbbell, for which $r_{_G}=l$ and Eq.(\ref{maximal-a-gyration})  always yields an unphysical impact parameter, so that the largest angular velocity is brought about by $a=l$ no matter what the dumbbell's mass. 

The  results that have just been reported  teach the lesson that one must refrain from very tempting but sometimes  hasty conclusions  even in elementary problems involving rotational motion of rigid bodies, for  intuition may be an easy prey to deception.

\end{document}